# Global Minimum Search via Annealing. Nanoscale Gold Clusters

Nadezhda A. Andreeva[1] and Vitaly V. Chaban[2,3,a]

1) PRAMO, Saint Petersburg, 197341, Russian Federation

2) Instituto de Ciência e Tecnologia, Universidade Federal de São Paulo, 12247-014, São José dos Campos, SP, Brazil

3) Department of Chemistry, University of Southern California, Los Angeles, CA 90089, United States

**Abstract**. Global minimum potential energy state can be very challenging to locate in a relatively large atomistic system. Our present work investigates this problem using an example of gold nanoclusters, $Au_{10}$, $Au_{20}$, $Au_{30}$, $Au_{50}$. Nanoscale gold particles (NGPs) contribute heavily life sciences through their applications in diagnostics and therapeutics. NGPs feature manifold atomistic configurations depending on the conditions of synthesis. We apply annealing molecular dynamics (AMD) as an alternative and supplement to the well-established eigenfollowing (EF) geometry optimization. We conclude that the combination of AMD and EF systematically works more efficiently than EF alone.

**Key words**: energy surface, annealing, nanoparticle, gold, band gap, PM7-MD.

---

[a] E-mail: vvchaban@gmail.com

TOC Image

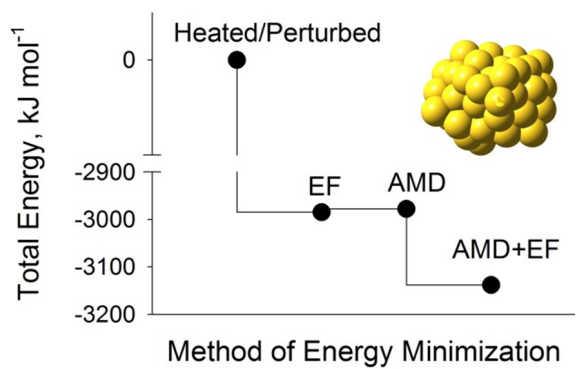

**Research Highlights**

(1) Potential energy surface of a set of gold nanoparticles was investigated.

(2) Annealing and eigenfollowing were used to obtain relaxed structures.

(3) More efficient scheme for global minimum search was introduced.

**Introduction**

Gold containing molecules and structures are omnipresent in pure chemistry and chemical technology.[1-11] Nanoscale gold particles (NGPs) are of significant importance and interest for modern chemistry, chemical engineering, and health sciences.[5,8,12,13] NGPs particles are employed in diagnostics and therapeutics.[14,15] Drastic alteration of chemical and spectroscopic properties of gold at the nanoscale is responsible for the success of such emerging applications.[15] Understanding of electronic structures of NG, as well as arrangement of gold atoms in relation to one another, contributes into the time- and cost-efficient design of NGPs. Experimental evidence exists that Au nanospheres, which are ca. 50 nm in diameter, exhibit a surface plasmon resonance peak at 520 nm.[14] At this resonant frequency, the incident light can be absorbed, being responsible for a ruby red color of these NGPs. Importantly, size and shape of NGPs play a major role in their optical properties. Further advance in engineering requires an ability to predict, control and characterize NGPs using both experimental manipulation techniques and chemical computational tools. The same considerations apply to nanostructures engendered by silver, platinum, palladium, etc.

It is a supposition following conventional wisdom that the real-world molecular and supramolecular structures attain global minimum configurations with respect to the applicable thermodynamic potential function. That is, the global minimum state is a function of external conditions, such as temperature and pressure. However, the larger the molecule, the more complicated energy surface (Figure 1) it supports due to a variety of covalent angles and dihedrals, plus abundant intramolecular non-bonded interactions. Such molecules may maintain their local minima states for a significant amount of time. The nanoparticles formed by *d*-elements constitute an ever more complicated case thanks to many allowed valence states. Consequently, investigation of large molecules and *d*-element structures is coupled with an epic problem to locate their most probable (i.e. most energetically favorable) atomistic arrangements. Another problem is to distinguish the actually implemented structure from other local minima structures exhibiting similar total energies. The former must depend on the conditions of preparation, since solid state formations are normally resistant to conformational changes.[16-18] After being synthesized by chemical means or self-assembled, principal alterations in the attained structure are not likely.

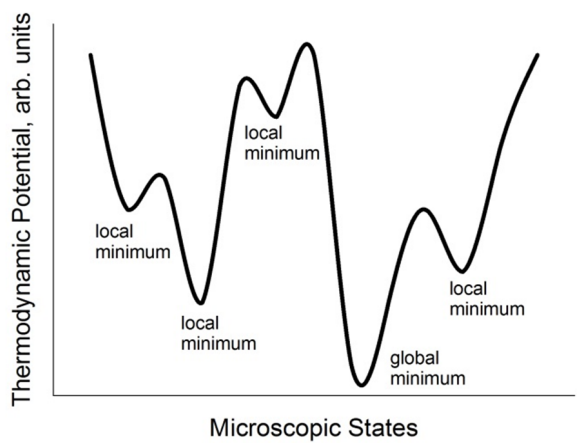

Figure 1. An exemplary potential energy surface featuring a plethora of local minima. Achievement of the global energy minimum state is challenging not only in numerical computer modeling, but also in the real-world techniques of chemical synthesis.

The recent review contributed by Jiang clearly highlights the difficulty to theoretically predict the structure of NGP of given composition at given conditions.[15] Only in a few cases the inherent structure of NGP was identified successfully, while the composition of plethora of NGPs can be relatively easily retrieved using mass spectrometry. The key challenge in this field facing the computational scientist is how to predict the equilibrium structures for known compositions, given a slow pace of experimental total structure determination. The problem is essentially global minimization of chemical structures, which is notorious for its principal complexity. According to Jiang,[15] density functional theory (DFT) has exhibited a remarkable success in the discussed research field. Mostly, pure DFT functionals within the generalized gradient approximation for electron exchange-correlation, have been a default choice for structure prediction.[19-21] The computational efficiency of pure DFT methods is somewhat higher than that of more accurate hybrid DFT methods. Over 11 years of efforts, the DFT energy was lowered by 9 eV. Finally, the correct structure was predicted in 2010,[22] shortly before the experimental structure was independently discovered.[23] Although the mentioned example is an obvious success story of DFT, these observations and considerations also underline the overall difficulty of structure determination, both experimentally and theoretically. While geometry optimization algorithms powered by DFT are able to predict a lot of possible structures, only one of them corresponds to the real-world chemistry. The computational speed of DFT is not high decreasing by approximately 2-3 orders of magnitude when the size of the system increases twice. Semiempirical methods are significantly faster, although accuracy of results and applicability of approximations must be carefully ensured.

This work reports usage of the eigenfollowing (EF) algorithm, which is one of the most popular geometry optimization schemes,[24] to obtain stable (local minimum) structures of $Au_{10}$, $Au_{20}$, $Au_{30}$, and $Au_{50}$ NGPs. We underline the limitations of EF. Instead of EF, we propose annealing molecular dynamics simulations (AMD) powered by the recently proposed PM7-MD method.[24,25] The proposed scheme appears more efficient in the global minimum search providing systematically lower energy structures than EF, while starting from the same point. The discrepancy between EF and AMD (in favor of AMD) increases as the size of NGP increases.

**Methodology**

Single-point semiempirical description[24,26-28] of nuclear-electronic system was applied to represent NGPs, $Au_{10}$, $Au_{20}$, $Au_{30}$, and $Au_{50}$ (Figure 2). The last parametrization of selected integrals by J.J.P. Stewart, PM7 ("parameterized model seven"), was used.[24] Total system energy, heat of formation, and forces acting on every Au nucleus were derived from the converged (self-consistence field criterion of $10^{-6}$ Hartree) PM7 wave function.

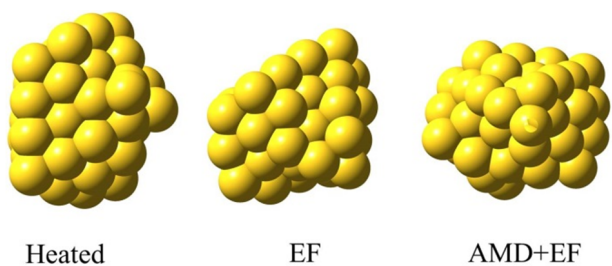

Figure 2. Structures of the Au$_{50}$ nanocluster: heated geometry (at 3000 K); geometry obtained via the eigenfollowing (EF) algorithm; geometry obtained via the combination of annealing molecular dynamics (AMD) and EF.

Coupled with canonical velocity distribution, the forces on every Au atom allow constructing Newtonian equations of motion. Equations of motion are propagated in time, with a time-step of 3.0 fs, using the Verlet algorithm. Both velocity and position are calculated at the same value of the time variable. The resulting consequence of atomic (nuclear) coordinates is often referred to as molecular dynamics (MD) trajectory. The conservation of total energy was preliminarily tested in the constant energy ensemble with respect to the applied integration time-step.

Since nuclear and electronic degrees of freedom evolve independently (no energy exchange between them is possible), the recorded trajectory obeys Born-Oppenheimer approximation. In the following, we will denote this simulation method PM7-MD.[25] Annealing PM7-MD simulations (AMD) were implemented using the Berendsen weak temperature coupling (with a relaxation constant of 50 fs).[29] The reference temperature of heat bath was changed uniformly by certain number of degrees Celsius at each integration time-step. The discussed setup of AMD effectively pulls kinetic energy away from the system. The system accommodates to the applied changes on-the-fly.

Eigenfollowing (EF) is a well-established geometry optimization algorithm, which experienced a great number of improvements (in particular, for stability) over the last decades.[24] It is based on the second order Taylor expansion of the energy around the current point. Energy and gradient are available at this point. Some estimation of the Hessian is additionally possible. This information is used to find the best step within or on the hypersphere with the current trust radius. The correct Hessian (if the search is performed for a minimum) must exhibit only positive eigenvalues. Then, a pure Newton-Raphson step is attempted. The step and the trust radius are updated. Finally, new energy and gradient are evaluated. The procedure is repeated until the requested geometry convergence criteria are met. EF is the default choice for geometry optimization in many quantum chemistry programs including MOPAC2012.[24]

The latest available revision of MOPAC2012[24] (obtained from Dr. J.J.P. Stewart) was used to obtain wave functions and forces. The Atomistic Simulation Environment (ASE) set of scripts[30] was used as a starting point to interface electronic structure stage of computations with temperature-coupled velocity Verlet trajectory propagation.

**Results and Discussion**

This section discusses the results of 19 AMD simulations and 19 EF geometry optimizations starting from various initial structures of $Au_{10}$, $Au_{20}$, $Au_{30}$, and $Au_{50}$ NGPs (Table 1). The initial structures of each composition were obtained via PM7-MD simulations at 3000 K (weak temperature coupling scheme) during 3-5 ps. The temperature of 3000 K is very well above the normal melting point of gold. Note that melting point of bulk gold and melting point of small NGPs are not identical. Such a high-temperature simulation can be considered an artificial perturbation of structure, which is still physical in the sense that energies and forces are obtained based on the Au structure. Seven initial structures of $Au_{10}$, five initial structures of $Au_{20}$, five initial structures of $Au_{30}$, and two initial structures of $Au_{50}$ were generated in such a way (Table 1). Each of these structures was optimized using EF. The heat of formation (Figure 3), HOMO-LUMO band gap (Figure 4), total energy (Figure 5), and net charges (Figure 6) were retrieved for these optimized structures. Simultaneously, these same initial structures were subject to AMD simulations, as detailed in Table 1. The NGPs obtained at the end of AMD simulations were additionally optimized using EF. This stage does not bring significant gain of formation energy, since the temperature of 300-500 K corresponds to solid state of NGPs. However, it is still important to remove minor geometry perturbations, which are the result of thermal motion. Remember, we apply AMD for the systems to freely wander over local minima and approach deeper ones. Final EF can conveniently deposit the system into the chosen state.

Table 1. Basic details of the simulated systems. Note that PM7 makes use of pseudopotentials to represent core electrons. Therefore, the provided number of electrons is much smaller than the total number of electrons in the corresponding NGPs. Faster cooling of smaller NGPs is possible because the reorientation dynamics in very small clusters (all Au atoms belong to the surface) is much faster as compared to larger structures

| System | # electrons | # MD runs | Annealing regime | Annealing time, ps | Perturbation time, ps |
|---|---|---|---|---|---|
| $Au_{10}$ | 110 | 7 | 1700-500 K | 10 | 3 |
| $Au_{20}$ | 220 | 5 | 1700-500 K | 15 | 3 |
| $Au_{30}$ | 330 | 5 | 1700-300 K | 25 | 5 |
| $Au_{50}$ | 550 | 2 | 1700-300 K | 40 | 5 |

Figures 3-4 compare energies and band gaps of structures obtained from geometry optimization and annealing PM7-MD simulations. The energies after AMD appear systematically smaller. Note the effect of system size. Four systems containing $Au_{10}$ exhibit nearly the same energies of the optimized structures despite the relaxation method used. However, the remaining three systems all exhibit more relaxed structures after AMD than after EF. One would expect that since $Au_{10}$ is a very small cluster (all atoms are located at the surface), both EF and AMD must provide the same output energy of formation. However, this is not the case and even AMD provides different stationary geometries, which differ by up to 20 kJ mol$^{-1}$. Abundance of stationary points at the potential energy surface of NGPs is, therefore, proven. None of the $Au_{20}$ and $Au_{30}$ systems obtained via EF is lower by energy than their analogues obtained via AMD. This observation is very important, since it largely explains why theoretically determined structures rarely match the experimental results. All equilibrium geometries of NGPs of appreciable size were, thus far, obtained via geometry optimization algorithms within the pure DFT methods. We anticipate that larger $Au_n$ (n>30) exhibit even stronger deviations if the EF algorithm is employed.

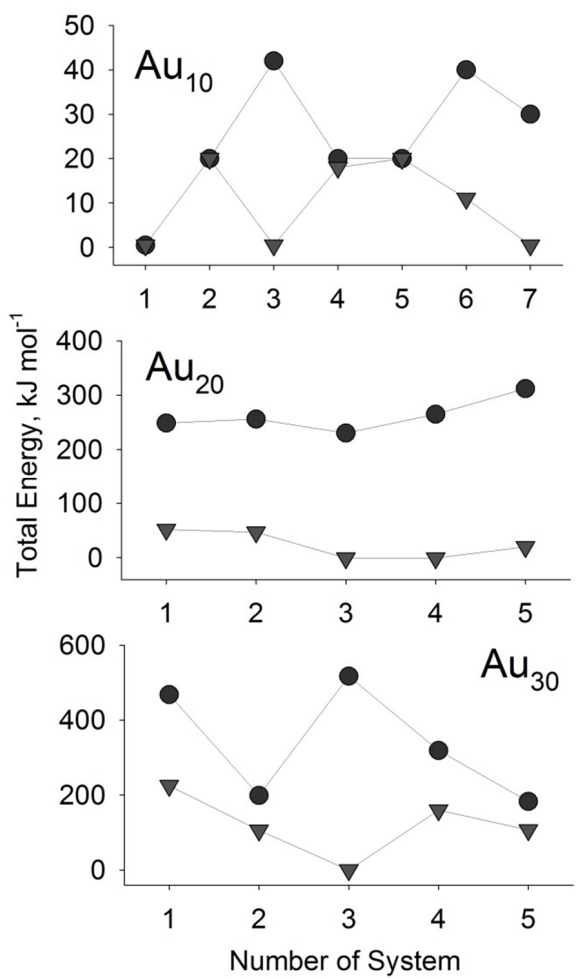

Figure 3. Energies (heats) of formation of the $Au_{10}$, $Au_{20}$, and $Au_{30}$ gold nanoclusters obtained via EF (circles) and via the combination of AMD and EF (triangles). The most energetically favorable nanocluster of each composition is assigned zero-energy for simplicity of representation. Therefore, all higher-energy nanoclusters exhibit relatively positive energies of formation.

The ultrasmall NGPs explored in this work possess large band gaps, which decrease as NGP increases in size (Figure 4). Remarkably, the band gap fluctuates significantly within the same chemical composition when EF is used. Fluctuations get smaller when the AMD+EF combination is employed. The band gap after AMD+EF is systematically smaller indicating smaller LUMO energy in these cases. Similar to energies of formation (Figure 3), band gaps underline the difference between EF and the AMD+EF combination. The observed difference is definitely in favor of AMD+EF.

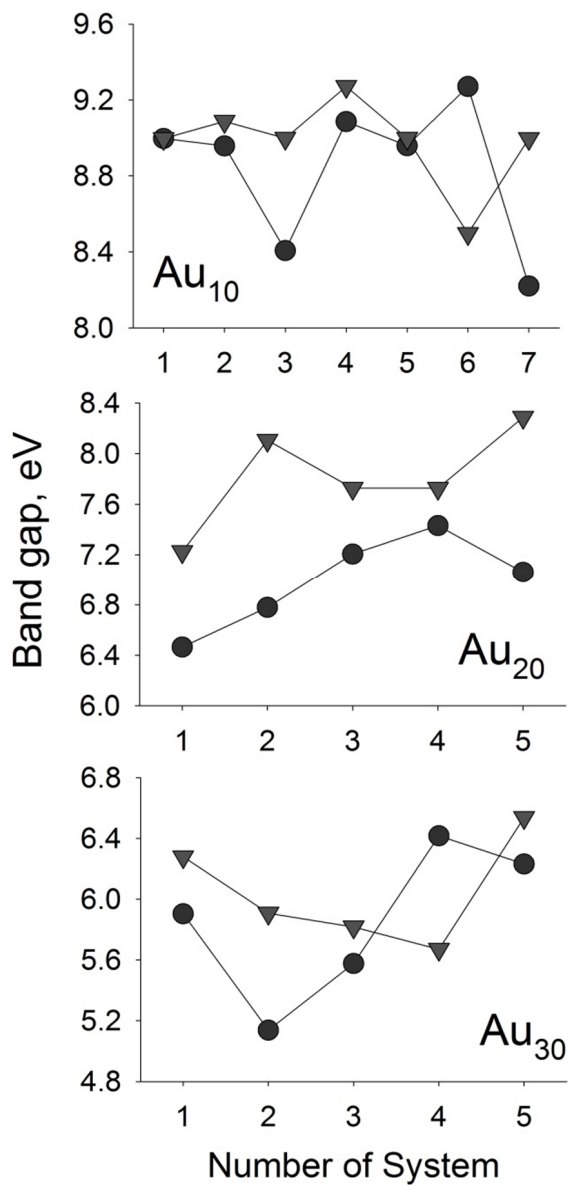

Figure 4. The band gaps of the $Au_{10}$, $Au_{20}$, and $Au_{30}$ gold nanoclusters obtained via EF (circles) and via the combination of AMD and EF (triangles). As expected, increase of the nanocluster size decreases a band gap.

Figure 5 summarizes the energetic differences between $Au_{50}$ (the largest NGP considered in the present work) prepared via various methods. The heated (perturbed) geometry after 5 ps of spontaneous molecular dynamics at 3000 K is obviously the least favorable structure. Application of EF decreases total nuclear-electron potential energy by 3000 kJ mol$^{-1}$. Continuous cooling of Au$_{50}$ from 1700 to 300 K during 40 ps provides nearly the same total energy decrease. However, when AMD was supplemented by EF, it resulted in the additionally total energy by ca. 150 kJ mol$^{-1}$. Ultimately, the combination of AMD+EF issues the most energetically favorable NGP structure. Note that energies in Figure 5 are averaged out using the data of the two independent systems containing Au$_{50}$.

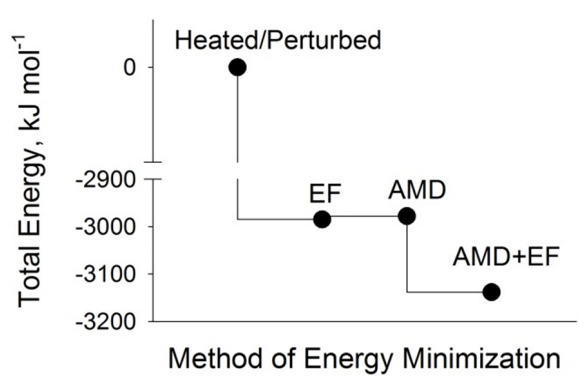

Figure 5. Total energy diagram demonstrating the difference between geometries of Au$_{50}$: obtained via high-temperature MD (heated/perturbed); optimized via the EF algorithm; obtained via AMD; obtained via the combination of AMD and EF. The depicted energies are averaged based on the two independent starting geometries of Au$_{50}$. All applicable methods and computational schemes are represented in details in the methodology section.

The net atomic charges in Figure 6 are defined based on the Coulson's charge assignment scheme (so-called Coulson charges). Similar to Mulliken charges, Coulson charges are derived from the population analysis of the wave function. These charges constitute an important measure of polarity of chemical bonds, which keep the molecule (nanostructure) as a whole. Bulk gold must theoretically exhibit zero charges on every atom, since they are chemically equivalent. NGPs constitute a different case, since the genuine symmetry of bulk gold is broken (Figure 6). Some atoms are electron-deficient, whereas others are electron-rich (ca. ±0.2 e). Significant portion of Au atoms is neutral or nearly neutral. As the structure gets optimized, i.e. total potential energy of the nuclear-electron system is decreased, the net charges also decrease.

However, a few significantly charged gold atoms remain unaltered (Figure 6, bottom). Population-based charges are an important measure of polar bonds, while polar bonds indicate chemical reactivity of the structure. Therefore, such an analysis, along with the HOMO-LUMO band gaps, provides important information about the obtained local minima spanning potential energy surface. Less polar covalent bonds between the Au atoms imply greater stability of the obtained NGP, including resistance to coalescence with other NGPs.

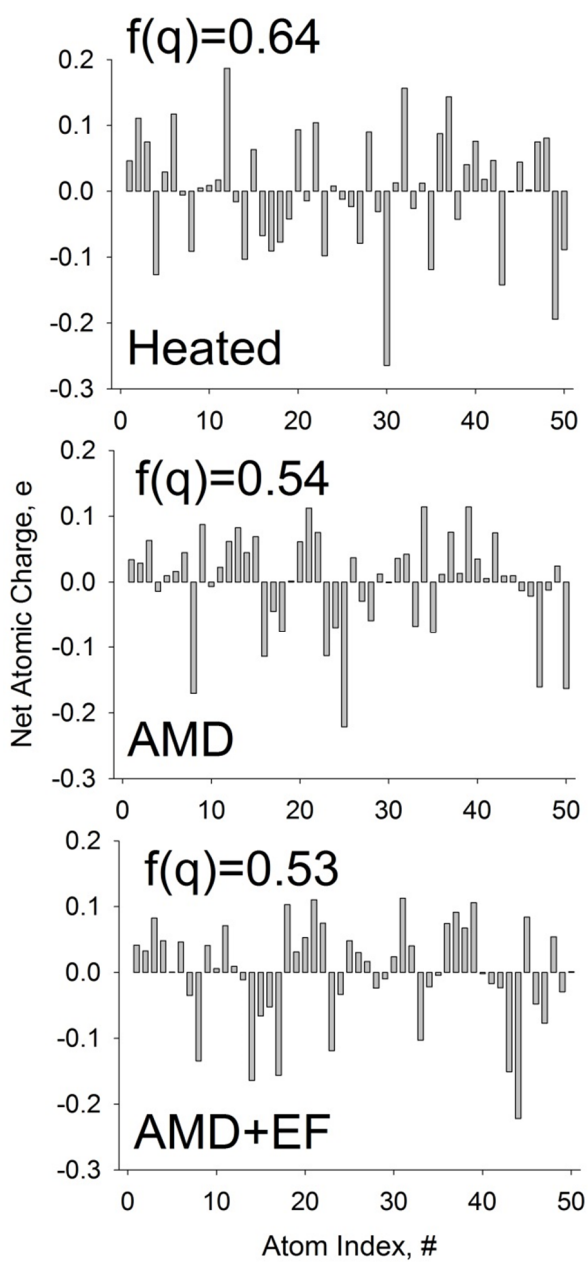

Figure 6. The net atomic charges computed in the three different states of the $Au_{50}$ nanocluster. See legends for details. Unlike in the bulk gold, a set of atomic charges in the nanocluster deviates from zero. To characterize deviation from zero, we introduce the following function,

$f(q_i) = \sqrt{\sum_i^N q_i^2}$ , where $q_i$ enumerates partial point charges on every Au atom. Note that better relaxed geometries feature smaller $f(q_i)$.

**Conclusions**

We have hereby investigated nanoscale gold nanoparticles of the following chemical composition, $Au_{10}$, $Au_{20}$, $Au_{30}$, and $Au_{50}$ using a semiempirical Hamiltonian, PM7. A variety of local minima states were obtained for each of the enumerated compositions. We illustrate that the NGP reaches different stationary states depending on the (1) starting nuclear configuration and (2) method of potential energy minimization. Local minimum states are abundant in even relatively small NGPs, such as $Au_{10}$. Their total number increases as the size of NGP increases. The eigenfollowing geometry optimization (i.e. structure energy minimization) method is only able to capture the closest local minimum with respect to the initial nuclear configuration of the NGP. If the initially supplied geometry is far from a global minimum, the resulting nuclear stationary state also appears far away meaning unnaturally high potential energy. Annealing molecular dynamics provides significant assistance in navigating over the potential energy surface of gold nanoparticles. According to our quantitative results, the combination of AMD and EF provides systematically lower energy structures than EF alone. The advantage of the newly proposed and evaluated computational scheme increases as NGP grows in size (number of atoms/electrons). The elaborated solution is, furthermore, beneficial for theoretical investigation of other nanoscale structures, such as those composed of silver, platinum, palladium, etc.

The recently introduced PM7-MD method[25] was selected as a potential energy calculator for annealing simulations, since it provides handy and transparent description of the *d*-element containing systems. Although classical additive potentials allow for faster calculations and better parallel scaling, their accuracy in the case of heavy elements is usually unsatisfactory, excluding most primitive structures and properties of interest.

**Acknowlegments**

V.V.C. is a recipient of the research grant from CAPES (Coordenação de Aperfeiçoamento de Pessoal de Nível Superior, Brasil) under "Science Without Borders" program. The idea and motivation of this work were inspired by multiple stays of V.V.C. in St. Petersburg in August, October, and November 2014.